\documentstyle[epsfig,floats,aps]{revtex}
\begin{document}
\draft
\twocolumn[\columnwidth\textwidth\csname@twocolumnfalse\endcsname
\title{Shell-model calculations of stellar weak interaction rates: \\
  II. Weak rates for nuclei in the mass range $A=45-65$ in supernovae
  environment}

\author{K. Langanke and G. Mart\'{\i}nez-Pinedo}

\address{Institut for Fysik og Astronomi, {\AA}rhus Universitet,
  DK-8000 {\AA}rhus C, Denmark}

\date{\today} 

\maketitle

\begin{abstract}
  Based on large-scale shell model calculations we have determined the
  electron capture, positron capture and beta-decay rates on more than
  100 nuclei in the mass range $A=45$-65. The rates are given for
  densities $\rho Y_e =10^7$-$10^{10}$~mol/cm$^3$ and temperatures
  $T=10^9$-$10^{10}$~K and hence are relevant for both types of
  supernovae (Type Ia and Type II). The shell model electron capture
  rates are significantly smaller than currently assumed. For
  proton-to-baryon ratios $Y_e=0.42$-0.46~mol/g, the beta-decay rates
  are faster than the electron capture rates during the core collapse
  of a massive star.
\end{abstract}

]

\section{Introduction}

Astrophysical environments can reach very high densities and
temperatures. Under these conditions (temperatures $T$ larger than a
few $10^9$ K), reactions mediated by the strong and electromagnetic
force are in chemical equilibrium and the matter composition is given
by nuclear statistical equilibrium \cite{Hoyle,Clayton}, i.e. it is
determined mainly due to the nuclear binding energies subject to the
constraint that the total number of protons in the composition
balances the number of electrons present in the environment.
Introducing the electron-to-baryon ratio $Y_e$ (units of mol/g), this
constraint can be formulated as
\begin{equation}
\sum_k \frac{Z_k}{A_k} X_k = Y_e;\quad \sum_k X_k = 1
\end{equation}
where the sum is over all nuclear species present and $Z_k$, $A_k$,
and $X_k$ are the proton number, mass number and mass fraction of
species $k$, respectively. Importantly, in these astrophysical
environments the relevant density and time scales are often such that
the neutrinos are radiated away, so that reactions mediated by the
weak interaction are not in equilibrium. Thus, weak interaction rates
play a decisive role in these environments changing $Y_e$ and hence
the composition of the matter.

Among these astrophysical environments are the two major contributors
to the element production in the universe: supernovae of type Ia and
type II (e.g.~\cite{Arnett}).  A type Ia supernova is usually
associated with a thermonuclear explosion on an accreting white dwarf.
Hydrogen mass flow from the companion star in the binary system at
rather high rates leads to steady hydrogen and helium burning on the
surface, increasing the carbon and oxygen mass of the white dwarf.
Finally carbon is ignited in the center of the star leading to a
thermonuclear runaway. A burning front then moves outwards through the
star at subsonic speed, finally leading to a detonation which explodes
the star. Several issues (including the masses of the stars in the
binary, the mass accretion history and composition, the matter
transport during the explosion, the speed of the burning front) are
currently still under debate (e.g.~\cite{Isern}).  It appears,
however, established that electron capture will occur in the burning
front driving the matter to larger neutron excess. As the observed
iron in the universe is made roughly in equal amounts by type Ia and
type II supernovae, type Ia supernovae should produce the relative
abundances within isotope chains in the iron mass region in agreement
within a factor of 2 with the observed solar
abundances~\cite{Iwamoto}.  Provided the electron capture rates are
sufficiently well known (matter is in nuclear statistical equilibrium
inside a type Ia supernova), this requirement allows to put severe
constraints on type Ia supernova models~\cite{Brachwitz}.

A type II supernova is related to the core collapse of a massive star.
Here the core of a massive star becomes dynamically unstable when it
exhausts its nuclear fuel. If the core mass exceeds the appropriate
Chandrasekhar mass, electron degeneracy pressure cannot longer
stabilize its center and it collapses. In the initial stage of the
collapse, electrons are captured by nuclei in the nickel mass range,
thus reducing $Y_e$. Associated is a decrease in degeneracy pressure
and energy, as the neutrinos can still leave the star; both effects
accelerate the collapse. With decreasing $Y_e$, i.e. with increasing
neutron excess of the nuclei present, $\beta$ decay becomes more
important and can compete with electron capture.

Under the stellar conditions discussed above, the weak interaction
rates are dominated by Gamow-Teller (GT) and, if applicable, by Fermi
transitions.  Bethe {\em et al.}~\cite{BBAL} recognized the importance
of the collective GT resonance for stellar electron capture.  Shortly
after, Fuller, Fowler and Newman (usually abbreviated as
FFN~\cite{FFN1,FFN2,FFN3,FFN4}) estimated the stellar electron capture
and beta-decay rates systematically for nuclei in the mass range
$A=45-60$ considering two distinct contributions. At first, these
authors estimated the GT contributions to the rates by a
parametrization based on the independent particle model. The rate
estimate has then been completed by including Fermi transitions and by
experimental data for discrete transitions, whenever available.
Unmeasured allowed GT transitions have been assigned an empirical
value ($\log ft=5$).  One of the important ideas in the seminal work
by FFN was to recognize the role played by the GT resonance in
$\beta^-$ decay via the GT back resonance in the parent nucleus (the
GT back resonance are the states reached by the strong GT transitions
in the inverse process (electron capture) built on the ground and
excited states, see~\cite{FFN2,FFN3}) allowing for a transition with a
large nuclear matrix element and increased phase space.  Until now the
FFN weak interaction rates are key ingredients in supernova
simulations.

The general formalism to calculate weak interaction rates for stellar
environment has been already given by Fuller {\em et
  al.}~\cite{FFN1,FFN2,FFN3,FFN4}. What had not been possible at the
time when FFN did their pioneering work was to solve the associated
nuclear structure problem with the necessary accuracy and predictive
power.  Several years ago, Aufderheide, Mathews and
collaborators~\cite{Aufderheide91,Aufderheide93a,Aufderheide93b}
pointed out that the interacting shell model is the method of choice
for this job. In fact, using the newly developed shell model Monte
Carlo techniques~\cite{Johnson,Koonin} electron capture rates for
selected nuclei have been derived~\cite{Dean98}, but it became
apparent that shell model diagonalization calculations are
preferable~\cite{Langanke98,Langanke99} as they allow for detailed
spectroscopy and do not have restrictions in their applicability to
odd-odd and odd-$A$ nuclei as the shell model Monte Carlo method has
at low temperatures~\cite{Koonin}. Before calculating weak interaction
rates for a large set of nuclei in the $A=45$-65 mass region, it had
to be proven that state-of-the-art shell model diagonalization is
indeed capable of reliably solving the required nuclear structure
problems.  This proof has been given in Ref.~\cite{Caurier99}. This
paper reported about large-scale shell model calculations covering the
relevant mass range. The studies were performed at a truncation level
in the $pf$ shell at which the GT strength distributions are virtually
converged. As residual interaction a slightly modified version of the
wellknown KB3 interaction~\cite{Zuker81} has been used; the slight
modifications correct for the small overbinding at the $N=28$ shell
closure encountered with the original KB3 force~\cite{Caurier99}.  In
general, it has been demonstrated that the shell model reproduces all
measured $GT_+$ distributions very well and gives a very reasonable
account of the experimentally known $GT_-$ distributions. Further, the
lifetimes of the nuclei and the spectroscopy at low energies is
simultaneously also described well.  Ref.~\cite{Caurier99} has
therefore shown that modern shell model approaches have the necessary
predictive power to reliably estimate stellar weak interaction rates.

In this paper we will use the shell model approach of~\cite{Caurier99}
and derive the weak interaction rates for more than 100 nuclei in the
mass range $A=45$-65 at a temperature and density regime relevant for
supernova applications. Our paper is organized as follows. In section
2 we will repeat the derivation of the necessary formalism for the
weak rates. The results are presented and explored in section 3. In
this section we will also compare them to the pioneering work of FFN.
In this comparison we will find systematic differences. The origin for
these differences will be explored and discussed in section 4.

\section{Stellar weak interaction rates formalism}

The definition of the stellar electron and positron capture and
$\beta$-decay rates has been derived by Fuller, Fowler and
Newman~\cite{FFN1,FFN2,FFN3,FFN4}. We will here  repeat the 
formulae and ideas which will allow us to explain our strategy and
procedure to calculate these rates and to point out differences with
previous compilations. 

\subsection{General formalism}

We computed rates for four processes mediated by the charged weak
interaction: 

\begin{mathletters}
  \label{eq:procdef}
  \begin{enumerate}
  \item Electron capture (ec),
    \begin{equation}
      \label{eq:ecdef}
      (Z,A) + e^- \rightarrow (Z-1,A) + \nu.
    \end{equation}
  \item $\beta^+$ decay ($\beta^+$),
    \begin{equation}
      \label{eq:b+def}
      (Z,A) \rightarrow (Z-1,A) + e^+ + \nu.
    \end{equation}
  \item Positron capture (pc),
    \begin{equation}
      \label{eq:pcdef}
      (Z,A) + e^+ \rightarrow (Z+1,A) + \bar{\nu}.
    \end{equation}
  \item $\beta^-$ decay ($\beta^-$),
    \begin{equation}
      \label{eq:b-def}
      (Z,A) \rightarrow (Z+1,A) + e^- + \bar{\nu}.
    \end{equation}
  \end{enumerate}
\end{mathletters}

The rate for these weak processes is given by

\begin{equation}
  \label{eq:rate}
  \lambda^\alpha = \frac{\ln 2}{K} \sum_i \frac{(2J_i+1)
    e^{-E_i/(kT)}}{G(Z,A,T)} \sum_j B_{ij} \Phi_{ij}^\alpha ,
\end{equation}
where the sums in $i$ and $j$ run over states in the parent and
daugther nuclei respectively and the superscript $\alpha$ stands for
ec, $\beta^+$, pc or $\beta^-$. The constant $K$ is defined as
\begin{equation}
K=\frac{2 \pi^3 (\ln 2) \hbar^7}{G_F^2 V_{ud}^2 g^2_V m_e^5 c^4},
\end{equation} 
where $G_F$ is the Fermi coupling constant, $V_{ud}$ is the up-down
element in the Cabibbo-Kobayashi-Maskawa quark-mixing matrix and
$g_V=1$ is the weak vector coupling constant.  $K$ can be determined
from superallowed Fermi transitions and we used $K=6146\pm
6$~s~\cite{Towner}.  $G(Z,A,T)=\sum_i \exp(-E_i/(kT))$ is the
partition function of the parent nucleus.  $B_{ij}$ is the reduced
transition probability of the nuclear transition. We will only
consider Fermi and GT contributions which, however, has been shown to
be quite sufficient:

\begin{equation}
  \label{eq:bij}
  B_{ij} = B_{ij}(F) + B_{ij}(GT).
\end{equation}
The GT matrix is given by:

\begin{equation}
  \label{eq:bgt}
  B_{ij}(GT) = \left(\frac{g_A}{g_V}\right)^2_{\text{eff}}
  \frac{\langle j||\sum_k \bbox{\sigma}^k \bbox{t}^k_\pm || i
  \rangle^2}{2 J_i +1},
\end{equation}
where the matrix element is reduced with respect to the spin operator
$\bbox{\sigma}$ only (Racah convention~\cite{edmons}) and the sum runs over
all nucleons.  For the
isospin rising and lowering operators, $\bbox{t}_\pm = (\bbox{\tau}_x
\pm i \bbox{\tau}_y)/2$, we use the convention $\bbox{t}_+ p = n$;
thus, `$+$' refers to electron capture and $\beta^+$ transitions and
`$-$' to positron capture and $\beta^-$ transitions.  Finally,
$(g_A/g_V)_{\text{eff}}$ is the effective ratio of axial and vector
coupling constants that takes into account the observed quenching of
the GT strength~\cite{osterfeld}. We
use~\cite{brownq,Langanke95,Martinez96}

\begin{equation}
  \label{eq:gaeff}
  \left(\frac{g_A}{g_V}\right)_{\text{eff}} = 0.74
  \left(\frac{g_A}{g_V}\right)_{\text{bare}},
\end{equation}
with $(g_A/g_V)_{\text{bare}} = -1.2599(25)$~\cite{Towner}. If the
parent nucleus (with isospin $T$) has a neutron excess, then the
GT$_-$ operator can connect to states with isospin $T-1$, $T$, $T+1$
in the daughter, while GT$_+$ can only reach states with $T+1$. This
isospin selection is one reason why the GT$_+$ strength is more
concentrated in the daughter nucleus (usually within a few MeV around
the centroid of the GT resonance), while the GT$_-$ is spread over
10-15~MeV in the daughter nucleus and is significantly more
structured.

The Fermi matrix element is given by:

\begin{equation}
  \label{eq:fermidef}
  B_{ij}(F) = \frac{\langle j || \sum_k \bbox{t}^k_\pm || i
  \rangle^2}{2 J_i +1}.
\end{equation}
In our calculations isospin is a good quantum number and the Fermi
transition strength is concentrated in the isobaric analog state (IAS)
of the parent state. Equation~(\ref{eq:fermidef}) reduces to,
\begin{equation}
  \label{eq:fermiT}
  B_{ij}(F) = T(T+1)-T_{z_i} T_{z_j},
\end{equation}
where $j$ denotes the IAS of the state $i$. We neglect the reduction
in the overlap between nuclear wave functions due to isospin mixing
which is estimated to be small ($\approx 0.5$\%~\cite{Towner}).

The last factor in equation~(\ref{eq:rate}), $\Phi_{ij}^\alpha$, is
the phase space integral given by

\begin{mathletters}
  \label{eq:phase}
  \begin{eqnarray}
    \label{eq:phase-ec}
    \Phi_{ij}^{ec} = \int_{w_l}^\infty w & p & (Q_{ij}+w)^2
    F(Z,w)\nonumber\\ 
    & & S_e(w) (1-S_\nu(Q_{ij}+w)) dw,
  \end{eqnarray}
  \begin{eqnarray}
    \label{eq:phase-b+}
    \Phi_{ij}^{\beta^+} = \int_1^{Q_{ij}} w & p & (Q_{ij}-w)^2
    F(-Z+1,w) \nonumber\\ 
    & & (1-S_p(w))(1-S_\nu(Q_{ij}-w)) dw,
  \end{eqnarray}
  \begin{eqnarray}
    \label{eq:phase-b-}
    \Phi_{ij}^{\beta^-} = \int_1^{Q_{ij}} w & p & (Q_{ij}-w)^2
    F(Z+1,w) \nonumber\\
    & & (1-S_e(w)) (1-S_\nu(Q_{ij}-w)) dw,
  \end{eqnarray}
  \begin{eqnarray}
    \label{eq:phase-pc}
    \Phi_{ij}^{pc} = \int_{w_l}^\infty w & p & (Q_{ij}+w)^2 F(-Z,w)
    \nonumber\\  
    & & S_p(w) (1-S_\nu(Q_{ij}+w)) dw,
  \end{eqnarray}
\end{mathletters}
where $w$ is the total, rest mass and kinetic, energy of the electron
or positron in units of $m_e c^2$, and $p=\sqrt{w^2-1}$ is the momentum
in units of $m_e c$. We have introduced the total energy available in
$\beta$-decay, $Q_{ij}$, in units of $m_e c^2$

\begin{equation}
  \label{eq:qn}
    Q_{ij} = \frac{1}{m_e c^2} (M_p - M_d + E_i -E_j),
\end{equation}
where $M_p, M_d$ are the nuclear masses of the parent and daughter
nucleus, respectively, while $E_i, E_j$ are the excitation energies of
the initial and final states. We have calculated the nuclear masses
from the tabulated atomic masses neglecting atomic binging energies.
$w_l$ is the capture threshold total energy, rest plus kinetic, in
units of $m_e c^2$ for positron (or electron) capture. Depending on
the value of $Q_{ij}$ in the corresponding electron (or positron)
emission one has $w_l=1$ if $Q_{ij} > -1$, or $w_l=|Q_{ij}|$ if
$Q_{ij} < -1$.  $S_e, S_p,$ and $S_\nu$ are the positron, electron,
and neutrino (or antineutrino) distribution functions, respectively.
For the stellar conditions we are interested in, electrons and
positrons are well described by Fermi-Dirac distributions, with
temperature $T$ and chemical potential $\mu$. For electrons,

\begin{equation}
  \label{eq:fermie}
  S_e = \frac{1}{\exp\left(\frac{E_e - \mu_e}{kT}\right)+1},
\end{equation}
with $E_e = w m_e c^2$. The positron distribution is defined
similarly with $\mu_p = - \mu_e$. The chemical potential, $\mu_e$, is
determined from the density inverting the relation

\begin{equation}
  \label{eq:inverme}
  \rho Y_e = \frac{1}{\pi^2 N_A}\left(\frac{m_e c}{\hbar}\right)^3  
\int_0^\infty (S_e-S_p) p^2 dp,
\end{equation}
where $N_A$ is Avagadro's number. Note that the density of
electron-positron pairs has been removed in~(\ref{eq:inverme}) by
forming the difference $S_e-S_p$.

In supernovae weak interactions with nuclei with mass numbers
$A=45$-65 occur at such densities that the neutrinos can leave the
star unhindered. Thus, there is no neutrino blocking of the phase
space, i.e. $S_\nu = 0$.

The remaining factor appearing in the phase space integrals is the
Fermi function, $F(Z,w)$, that corrects the phase space integral for
the Coulomb distortion of the electron or positron wave function near
the nucleus.  It can be approximated by
\begin{equation}
  \label{eq:Fzw}
  F(Z,w)=2(1+\gamma)(2pR)^{-2(1-\gamma)} \frac{|\Gamma (\gamma +
    i y)|^2}{|\Gamma (2\gamma+1)|^2} e^{\pi y},
\end{equation}
where $\gamma=\sqrt{1-(\alpha Z)^2}$, $y=\alpha Zw/p$, $\alpha$ is the
fine structure constant, and $R$ is the nuclear radius.

Finally, the calculation of the rates reduces to the evaluation of the
nuclear transition matrix elements for the GT operator. The problem
obviously lies in the fact that many states (can) contribute to the
two sums over $i,j$. At first, the finite temperature allows the
thermal population of excited states in the parent. Each of these
states is then connected to many levels in the daughter nucleus by the
GT operators. A state-by-state evaluation of both sums is still beyond
present-day computer abilities. Before we summarize our strategy to
approximate the sums we recall that previous compilations of the
stellar weak rates employed the so-called Brink hypothesis
(e.g.~\cite{Aufderheide94}): Let $S_0 (E)$ be the GT distribution in the
daughter nucleus build on the ground state, then it is assumed that
the distribution $S_i (E)$ build on the excited state in the parent at
excitation energy $E_i$ is the same as $S_0 $, but shifted in energy
by $E_i$, i.e. $S_0(E)=S_i (E+E_i)$. This hypothesis has been tested
in various shell model calculations and is found to be valid for the
gross structure of the GT distribution.  However, it can be badly
violated for specific transitions as they occur at low excitation
energies (and are important for nuclear lifetimes). This tells us that
Brink's hypothesis should not be employed if specific low-lying
transitions (which usually exhaust a very small fraction of the total
GT strength) dominate the rates or are important. Such situations may
occur at low temperatures and densities. When the temperatures and
densities are higher so that many states (a few tens or more)
contribute, variations in low-lying transition strengths tend to
cancel and Brink's hypothesis becomes a valid approximation. We will
demonstrate this in detail later in this paper.

\subsection{Evaluation of the rates and checks}

Our strategy to calculate the weak rates~(\ref{eq:procdef}) is best
explained by considering a pair of nuclei $(Z,A)$ and $(Z+1,A)$
connected by the weak processes under discussion here. We have then
calculated the GT$_+$ distributions for all individual levels in the
nucleus $(Z+1,A)$ at modest excitation energies and the GT$_-$
strength distributions for the low-lying levels in the nucleus
$(Z,A)$.  Whenever experimental information about excitation energies
or GT transition strengths is available, the shell model results have
been replaced by data.  Otherwise the shell model predictions are
used. The quality of these calculations is demonstrated
in~\cite{Caurier99} where calculated spectra and lifetimes are
compared to data. For even-even parents we considered explicitly the
lowest $0^+$, $4^+$ and the two lowest $2^+$ levels which describe the
spectrum typically upto excitation energies of 2~MeV.  Depending on
the level spectrum, we adopted between 4 and 12 individual states for
odd-$A$ and odd-odd parent nuclei including at least the shell model
spectrum at excitation energies below 1 MeV explicitly. As our shell
model GT distributions have been calculated with 33 Lanczos iterations
(see \cite{Caurier99}), a state in the parent nucleus is connected to
33 ( for angular momenta $J=0$) til 99 (for angular momenta $ J\geq
1$) states in the daughter nucleus by the GT$_+$ operator, while this
magnifold is tripled for the GT$_-$ operator due to the different
isospin final states. (However, the T+1 component does not play a role
in calculating the rates under the temperature/density conditions we
are concerned with and is often omitted.)

While the explicit consideration of the low-lying states guarantees a
reliable description of the rates at low temperatures/densities, where
individual transitions are often decisive, states at higher excitation
energies become increasingly important at the higher temperature and
density regimes under consideration here. This is particularly true
for the $\beta^-$ decay rates which is often dominated by the
back-resonances under these conditions~\cite{FFN2,FFN3}. We have
therefore supplemented the contribution of the low-lyin states by the
back-resonance contributions which can be derived from the low-lying
contributions in the inverse process, i.e. the states populated by
electron capture on ($\beta^-$ decay of) low-lying states became the
back-resonances in the $\beta^-$-decay (electron capture). The
back-resonance contribution defined this way does not exhaust the total
GT strength built on this excited state. In particular, 
the capture of high energy electrons in the
back-resonance states will lead to states in the daughter which are
not included in our model. To correct for these missing transitions we
have employed the Brink hypothesis, i.e. we calculate the total GT
strength and the centroid $E_c$ for the parent ground state and place
the strength of the back-resonance state at an energy $E_i+E_c$, where
$E_i$ is the energy of the back-resonance state in the parent.

\begin{figure}[ht]
  \begin{center}
    \leavevmode
    \epsfig{file=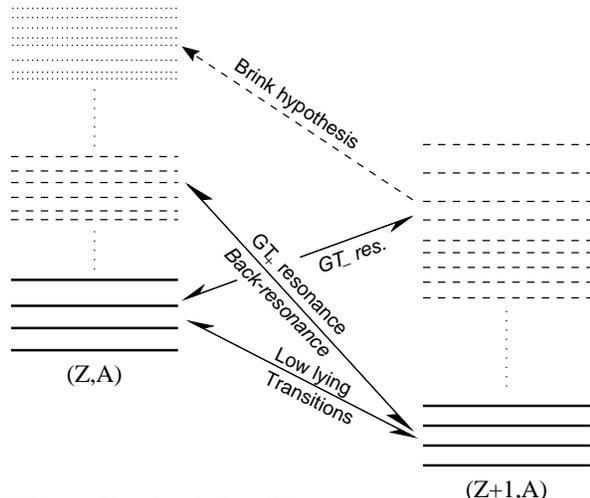,width=0.9\columnwidth}
    \caption{Sketch of the GT transitions considered in the present
      estimate of the stellar weak interaction rates. Our evaluation
      of the sums in Eq.~\ref{eq:rate} explicitly includes the GT$_-$
      and GT$_+$ strength distributions for the few lowest states in a
      nucleus. This is supplemented by the back-resonances which are
      derived from the GT strength distributions of the inverse
      process, as described in the text. Experimental data for the
      energies and GT transition strengths have been used, whenever
      available. Otherwise the shell model results are adopted.
      Finally the Brink hypothesis (see text) is employed to derive
      the remaining GT strength and centroid position for
      back-resonance states. In the sketch these transitions are shown
      by dashed lines, while the GT transitions, for which either data
      or shell model results are used, are drawn by solid lines.}
    \label{fig:schema}
  \end{center}
\end{figure}

Fig.~\ref{fig:schema} illustrates the various contributions to the two
sums in the rate formulae. In evaluating these sums we typically
consider several hundred states in both the parent and the daughter
nucleus.  The partition function is consistently derived from the same
parent states. We do not introduce a cut-off of levels at the particle
separation thresholds~\cite{BBAL,Tubbs}.

To demonstrate the convergence of our rates we will explicitly discuss
the pair of nuclei $^{63}$Co and $^{63}$Ni. Due to
Ref.~\cite{Aufderheide94}, $^{63}$Co is among the most important
$\beta$-decaying nuclei in a presupernova collapse at densities around
$10^8$ g/cm$^3$, while electron capture on $^{63}$Ni plays a moderate
role at the same astrophysical conditions.  The $Q_\beta$ value for
this pair of nuclei is 3.672(20)~MeV. Thus under laboratory conditions
$^{63}$Co $\beta$-decays to $^{63}$Ni, leading dominantly to the
excited $5/2$ state at 87~keV with an $\log ft$ value of 4.8(1).  Our
shell model calculation agrees with the data.

Our calculation of the stellar electron capture (and $\beta^+$ decay)
rate explicitly considers the two lowest $1/2^-$ and $3/2^-$ states
and the lowest $5/2^-$ state in $^{63}$Ni, which comprises the
experimentally known spectrum upto an excitation energy of 1~MeV.
Furthermore we include the back-resonances from the inverse reaction
which introduces a total of 6 states in the excitation energy interval
between 1 and 2 MeV, while experimentally 8 states are known in this
energy regime. Between 2 and 5 MeV, we include 16 more states which,
however, are probably not fully converged within our Lanczos procedure
and thus represent `averaged GT states' rather than physical states.

Table~\ref{tab:convec} demonstrates the convergence of the stellar
$^{63}$Ni electron capture rate with the number of initial states. We
have calculated the rates for various densities between $\rho
Y_e=10^7$ and $10^{10}$ mol/cm$^3$. The temperatures have been chosen
from the FFN temperature/density grid to be the closest to the
expected stellar trajectory in a supernova
collapse~\cite{Aufderheide94}. The calculations have been performed
considering only the ground state, the lowest 2 and 3 states, all 5
states for which shell model $GT_+$ distributions have been explicitly
calculated and finally for our full procedure including individual
states and back-resonances.

The calculation performed by restricting the sum over initial states to
only the ground state resembles the application of `Brink's shift
hypothesis', as with this assumption the nuclear matrix elements and
the phase space factors loose their dependence on the parent state and
the sum over initial states cancels the partition
function~\cite{Aufderheide94}.  (We point out that this calculation
is, however, not the same as those in
Refs.~\cite{FFN2,FFN3,Aufderheide94} as we adopt the shell model
GT$_+$ strength distribution, while the other authors had used an
empirical parametrization of the GT strength (see below).)  As
expected, the application of the Brink hypothesis is not a good
approximation at low temperatures and densities where the rate is
sensitive to low-lying transitions which can vary strongly between
various initial states.  The assumption becomes, however, quite
acceptable at higher densities where enough high-energy electrons are
available to effectively reach the centroid of the GT distribution in
the daughter. As stated above, Brink's shift hypothesis works well if
centroids are compared. To demonstrate this quantitatively we have
calculated the GT centroids $(E_{GT}$) for the 5 initial states. A
measure for the validity of the Brink hypothesis is then given by the
difference $E_{GT}-E_x$, where $E_x$ is the shell model excitation
energy of the initial state. If measured relative to the $^{63}$Ni
ground state, this difference is 13.7 MeV for the ground state, while
we find 13.7 MeV ($5/2^-$), 13.7 MeV ($3/2^-$), 13.9 MeV (excited
$3/2^-$) and 14.1 MeV (excited $1/2^-$) for the other states.

\begin{table*}[ht]
  \caption{Convergence of the stellar $^{63}$Ni electron capture rate,
    $\log \lambda^{\text{ec}}$, along the stellar trajectory during a
    supernova  collapse (defined by temperature $T_9$ and density
    $\rho$). Our full rate is compared to rates in which only the
    lowest 5, 3, 2, and 1 states in $^{63}$Ni are considered. The `no
    Brink' calculation has been performed neglecting those rate
    contributions which are shown by dashed arrow and are labelled
    `Brink hypothesis' in Fig.~\ref{fig:schema}.}
  \label{tab:convec}
  \begin{center}
    \begin{tabular}{ccdddddd}
      $T_9$ & $\log \rho Y_e$ & full & no
      Brink &5 states & 3 states & 2 states & 1 state  \\
      \hline
      3 &  7 & $-$8.46 & $-$8.46 & $-$8.62 & $-$8.59 & $-$8.46 &
      $-$9.62 \\
      4 &  8 & $-$5.18 & $-$5.18 & $-$5.29 & $-$5.25 & $-$5.12 &
      $-$5.83 \\ 
      5 &  9 & $-$1.74 & $-$1.77 & $-$1.77 & $-$1.72 & $-$1.61 &
      $-$1.93 \\ 
      7 & 10 &    1.78 &    1.63 &    1.74 &    1.78 &    1.80 &
        1.84 
    \end{tabular}
  \end{center}
\end{table*}

In Table~\ref{tab:convec} we also observe that, even at low densities,
a convergence is rather fast achieved if the rate comprises a thermal
average over a few states.

The $^{63}$Co $\beta$-decay rate has been calculated by explicitly
considering the GT$_-$ distributions for the lowest $7/2^-$ (ground
state), $3/2^-$ (at 0.995~MeV), $5/2^-$ (at 1.437 MeV) and $1/2^-$ (at
1.889 MeV) states. Only the $1/2^-$ state has a definitive spin
assignment, the spin for the other states has been assigned based on
our calculations. These contributions are then supplemented by the
back-resonances obtained from the GT$_+$ distributions of the inverse
reaction. These include another 4 states in the excitation energy
interval between 1 and 2~MeV and 18 (average) states between 2 and 5
MeV. Due to their construction, the back-resonances have only
transitions to the 5 lowest levels in $^{63}$Ni. The missing decay
channels of these states are, however, significantly suppressed due to
the strong $Q_\beta$ dependence.

In Table~\ref{tab:convbd} we study the convergence of the $^{63}$Co
$\beta^-$ decay rate.  The full rate is compared to rates in which the
contributions from the 4 individual levels are kept, but the
back-resonance contributions are cut at 5 MeV and 3 MeV in the parent
nucleus.  Additionally we have calculated the rates only from the
GT$_-$ distributions of the lowest 4, 2, and 1 states, totally
neglecting the back-resonances.

We find that at low temperature/densities the rate can be calculated
solely from the individual levels in a good approximation; even the
decay rate determined from the ground state alone is already a fair
approximation. At moderate and higher densities the back-resonances
become increasingly more important. This is easily understood by the
fact that at these high densities the Fermi energy of the electrons
gets so high that decays from low-lying states are effectively
blocked.  Correspondingly the $\beta$ decay rates decrease with
increasing density.  But we also note from 
Table~\ref{tab:convbd} 
that the rate still
converges rather rapidly and that considering the back-resonances upto
5 MeV excitation energy gives a sufficient approximation.

\begin{table*}[ht]
  \caption{Convergence of the stellar $^{63}$Co $\beta^-$ decay rate,
    $\log \lambda^{\beta^-}$, along the stellar trajectory during a
    supernova collapse (defined by temperature $T_9$ and density
    $\rho_7$). Our full rate is compared to rates in which
    back-resonant states are only considered upto 5 MeV and 3 MeV,
    respectively, and to rates which are calculated from the GT$_-$
    distributions of the lowest 4, 2, and 1 states in $^{63}$Co.}
  \label{tab:convbd}
  \begin{center}
    \begin{tabular}{ccdddddd}
      $T_9$ & $\log \rho Y_e$ & full & cut 5 MeV & cut 3 MeV & 4
      states & 2 states & 1 state  \\  
      \hline
      3 &  7 & $-$1.60 & $-$1.60 & $-$1.60 & $-$1.61 & $-$1.62 & $-$1.64 \\
      4 &  8 & $-$1.77 & $-$1.77 & $-$1.79 & $-$1.89 & $-$1.90 & $-$1.97 \\
      5 &  9 & $-$2.55 & $-$2.55 & $-$2.71 & $-$3.52 & $-$3.65 & $-$4.03 \\
      7 & 10 & $-$5.43 & $-$5.51 & $-$5.99 & $-$7.12 & $-$7.30 & $-$7.71
    \end{tabular}
  \end{center}
\end{table*}

We mention that, due to~\cite{Aufderheide94}, at high densities nuclei
in the supernova environment are expected to be more neutron-rich than
$^{63}$Co. Thus the relevant $Q_\beta$ values for $\beta$ decay
increase making the final-state blocking by electrons less effective
and reducing the importance of back-resonant states relative to the
low-lying individual states which we have considered.

\section{Stellar weak rates}

We have calculated the stellar weak interaction rates (electron and
positron capture, $\beta^-$ and $\beta^+$ decay) for more than 100
nuclei in the mass range $A=45$-65. The rates have been calculated 
for the same temperature and density grid as the standard FFN
compilations~\cite{FFN2,FFN3,FFNNET}. An electronic table of our rates
is available from the present authors upon request. We have also
prepared a table in which the electron capture rates are presented in
terms of the `effective rates' introduced and defined by Fuller {\em
  et al.\/}~\cite{FFN4} which allow a more reliable interpolation.

\begin{figure}[ht]
  \begin{center}
    \leavevmode
    \epsfig{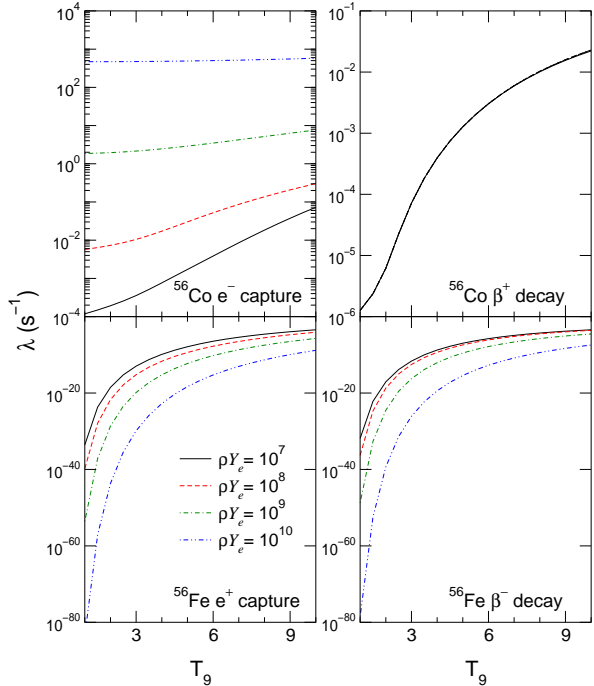}
    \caption{Stellar weak interation rates for the pair of nuclei
    $^{56}$Fe and $^{56}$Co as a function of $T_9$ for selected values
      of $\rho Y_e$ (in mol cm$^{-3}$)}
    \label{fig:coferat}
  \end{center}
\end{figure}

\begin{figure}[ht]
  \begin{center}
    \leavevmode
    \epsfig{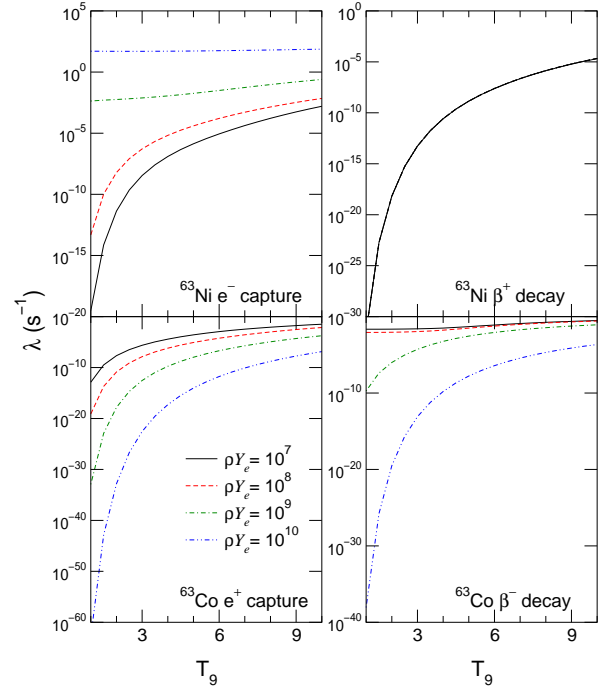}
    \caption{Stellar weak interation rates for the pair of nuclei
      $^{63}$Co and $^{63}$Ni as a function of $T_9$ for selected
      values of $\rho Y_e$ (in mol cm$^{-3}$)}
    \label{fig:conirat}
  \end{center}
\end{figure}

Examples of our rates are shown in figures~\ref{fig:coferat}
and~\ref{fig:conirat} using the pairs of nuclei ($^{63}$Co, $^{63}$Ni)
and ($^{56}$Fe,$^{56}$Co) as typical examples. The latter pair
includes an even-even and odd-odd nucleus. The $Q_\beta$ value for
these nuclei is negative and in the laboratory $^{56}$Co decays to
$^{56}$Fe by electron capture. To calculate the electron capture rate
we have considered the 5 states in $^{56}$Co below 1 MeV excitation
energy and the lowest $1^+$ state (at 1.72 MeV) explicitly,
supplemented by the back-resonances.  The $^{56}$Fe $\beta$ decay rate
can be calculated from the back-resonances due to the negative $Q$
value.

The electron capture rates on both nuclei $^{63}$Ni and $^{56}$Co
increase with temperature and density. Due to the negative $Q_\beta$
value, electron capture is possible from all states in $^{56}$Co and
thus the increase with temperature is rather mild. However, in the
case of $^{63}$Ni electron capture is quite sensitive to temperature (at low
densities), as it has to overcome a threshold of
nearly 4~MeV which at the low densities ($\rho Y_e=10^7$ and $10^{10}$
g/cm$^3$) is
mainly achieved by thermal population of excited states in the parent
nucleus, as the Fermi energies of the electrons (1.2~MeV and 2.4~MeV,
respectively for $T_9=1$) is still noticeably smaller than the
$Q_\beta$ value of 3.67~MeV. However, at the higher densities
electrons with energies above $Q_\beta$ are sufficiently available to
allow for efficient capture.  In this case, the rate becomes only
slightly dependent on temperature.

The $^{56}$Fe $\beta$-decay rate shows a very steep temperature
dependence and decreases with density. Both effects are readily
explained.  Due to the threshold of about 4.5 MeV, $\beta$ decay is
only possible from moderately excited states which have to be
populated thermally leading to the strong temperature dependence. As
the electron Fermi energy increases with density, an increasingly
larger part of the phase space gets Pauli-blocked. Consequently the
$\beta$-decay rate decreases with increasing density. Note that the
centroid of the GT$_+$ strength distributions of the $^{56}$Co ground
state is at an excitation energy of around 8 MeV in $^{56}$Fe. Thus at
densities larger than $\rho Y_e =10^9$~mol/cm$^3$ (corresponding to an
electron Fermi energy of about 5~MeV for $T_9=5$), the strong
transitions associated with the centroid of the back-resonances are
getting Pauli-blocked explaining the larger decrease with density even
at high temperatures.

$^{63}$Co can $\beta$ decay from all states and indeed already the
ground state has a strong GT transition to the excited state in
$^{63}$Ni at 87 keV. Furthermore, the first excited state is at nearly
1 MeV excitation energy. Consequently, the $^{63}$Co $\beta$-decay
rate shows only a mild temperature dependence at low densities.
Pauli-blocking by the electrons in the final state becomes, however,
important at higher densities introducing effectively a threshold for
the $\beta$ decay which has to be overcome by thermal population of
excited states in the parent. As a result, the rate develops an
increasing sensitivity to temperature. The centroid of the GT$_+$
strength distribution of the $^{63}$Ni ground state is at an
excitation energy of around 2.5 MeV in $^{63}$Co. Consequently Pauli
blocking by electrons becomes increasingly important for densities
above $10^9$ g/cm$^3$.

The positron distribution does not play a role for the $\beta^+$ decay
rate, which is virtually independent on density for both nuclei
$^{63}$Ni and $^{56}$Co (this is already apparent from the FFN
rates~\cite{FFN2,FFN3}).  The $^{56}$Co $\beta^+$ rate depends only
mildly on temperature. This is different for $^{63}$Ni where a
threshold has to be overcome by thermal population in the parent. For
both nuclei the $\beta^+$ decay rate is noticeably smaller than the
electron capture rate which generally dominates the weak rates for
charge-decreasing nuclear transitions under supernova conditions.

The positron capture rate decreases with density, but increases with
temperature. Both dependencies are caused by the positron distribution
where an increasing number of high-energy positrons gets available by
raising the temperature or lowering the density. The latter is caused
by the fact that the degeneracy parameter $\mu/kT$ is negative for
positrons. Due to its threshold for positron capture, the rate on
$^{56}$Fe shows the steeper temperature dependence. Again, the
positron capture rates are usually smaller than the competing
$\beta^-$ rates under supernova conditions for the nuclei of interest
here.

\begin{figure}[ht]
  \begin{center}
    \leavevmode
    \epsfig{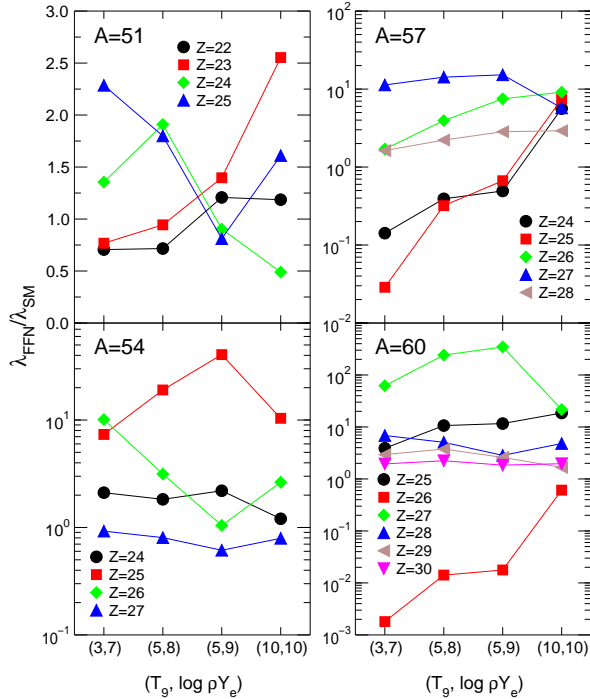}
    \caption{Ratio of the FFN and shell model electron capture rates
      for nuclei in the mass chains $A=51$, 54, 57, and 60 for
      selected values of temperature and density, $T$ and density,
      $\log(\rho Y_e)$. The charge numbers refer to the parent
      nucleus.}
    \label{fig:ec-ffn-sm}
  \end{center}
\end{figure}

\begin{figure}[ht]
  \begin{center}
    \leavevmode
    \epsfig{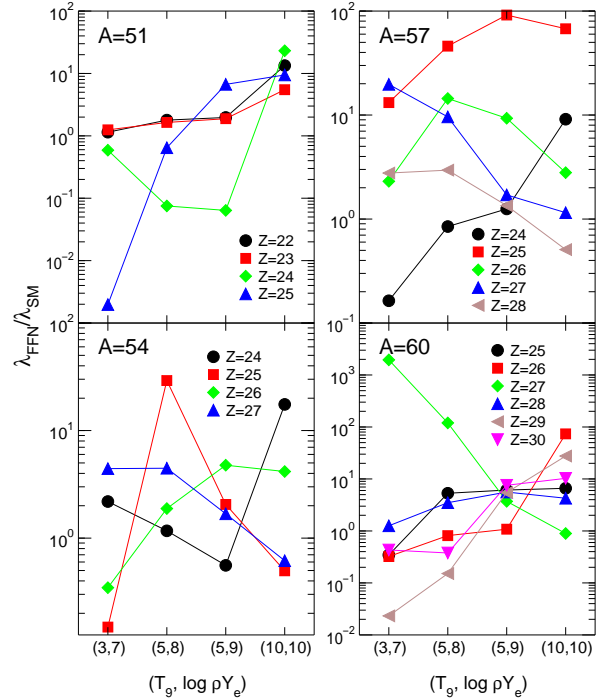}
    \caption{Ratio of the FFN and shell model $\beta$-decay rates for
      nuclei in the mass chains $A=51$, 54, 57, and 60 for selected
      values of temperature $T$ and density, $\log(\rho Y_e)$. The
      charge numbers refer to the daughter nuclei.}
    \label{fig:bd-ffn-sm}
  \end{center}
\end{figure}

The most interesting question clearly is: How do the shell model rates
compare to the FFN rates?

To answer this question, we have calculated the ratio
$\lambda_{\text{FFN}}/\lambda_{\text{SM}}$, where
$\lambda_{\text{FFN}}$ and $\lambda_{\text{SM}}$ are the FFN and shell
model rates, respectively. The FFN rates are taken from the electronic
file available at~\cite{FFNNET}. For the comparison we choose 4
different temperature and density grid points $(T_9, \log (\rho
Y_e)$): (3,7), (5,8), (5,9) and (10,10). Ratios for the two important
weak processes, electron capture and $\beta^-$ decay, are plotted in
figures~\ref{fig:ec-ffn-sm} and~\ref{fig:bd-ffn-sm} for 4 different
chains of isotones equally spanning the mass range between $A=50$-60.

To understand the differences observed in figures~\ref{fig:ec-ffn-sm}
and~\ref{fig:bd-ffn-sm} we have to recall how the FFN rates have been
derived. At first, these authors considered experimental data for
discrete transitions, whenever available (like we do in the present
shell model rates). The main contribution to the FFN rates usually
comes from the so-called GT resonance which they parametrized on the
basis of the independent particle model and which represents the total
GT strength by a single state. Here it is not so important that the
authors did not explicitly consider the quenching of the GT strength
with respect to the independent particle model\footnote{The quenching
  in the GT$_+$ strength had not been established at the time FFN
  calculated the rates. In later discussions~\cite{FFN4} these authors
  point to this effect and showed a way how to incorporate quenching
  effectively into the rates.}.  More relevant is where FFN placed the
GT resonance.  Here we will concentrate on the electron capture. The
differences in the $\beta^-$ decay rates follow then from a discussion
of the back-resonance contributions.

FFN estimated the GT resonance energy $E_{GTR}$ from 3 distinct
contributions:
\begin{equation}
E_{GTR} = \Delta E_{sp} + \Delta E_{ph} + \Delta E_{pair}.
\end{equation}
The single particle term is calculated as follows.  Starting with the
independent particle model wave functions for the ground states of the
parent and daughter nucleus, protons are acted on with the GT$_+$
operator leading to final neutron states, which, within the independent
particle model, correspond to 1p-1h excitations of the ground state
(or is the ground state). The corresponding excitation energy is
readily calculated, where FFN used the single particle energies of
Seeger~\cite{Seeger}. If the GT$_+$ operation can lead to several
different final states the excitation energy of the resonance is
computed taking the weighted average of the different transitions.
$\Delta E_{ph}$ is the particle-hole repulsion energy which has to be
supplied to pull a neutron out of the daughter ground state.  For
simplicity, FFN put $\Delta E_{ph} = 2$ MeV for all nuclei. Finally,
FFN argued that there is a penalty energy which has to be paid to
break a neutron pair if there is an even number of neutrons in the
daughter ground state. When applicable, this pairing energy is
approximated by 2 MeV for all nuclei.

In previous publications~\cite{Martinez99} we have already pointed out
that the shell model rates for important electron-capturing nuclei
along the stellar trajectory (due to the ranking given in
\cite{Aufderheide94} these are generally odd-$A$ and odd-odd nuclei)
are usually smaller than the FFN rates. As one potential reason for
this difference we have noted that the GT centroids for odd-$A$ and
odd-odd nuclei are usually at higher energies than assumed by FFN. We
will discuss this difference and its potential origin in details
below, but we note here that there are two other important ingredients
(low lying transitions and $Q_\beta$ values) which can lead to
differences between the shell model and FFN electron capture rates.
In fact, for $^{51}$Ti and $^{51}$V we find from
fig.~\ref{fig:ec-ffn-sm} that the shell model rates are larger than
the FFN rates at low temperature/density. For these cases the
difference is due to the fact that the shell model predicts low-lying
strength for the ground states (in $^{51}$V to the second excited
$5/2$ state in $^{51}$Ti, in $^{51}$Ti to the lowest $1/2$ state in
$^{51}$Sc) which are larger than the standard assignment ($\log ft=5$)
used in~\cite{FFN1,FFN2,FFN3} for experimentally not known
transitions. At larger temperatures/densities the low-lying strengths
becomes less important as the capture proceeds mainly to the GT
resonance; consequently the FFN rates are then larger than the shell
model rates due to differences in the position of the centroid. (Note
that the GT$_+$ strength distribution for $^{51}$V has been measured
and it agrees nicely with the shell model results~\cite{Caurier99}.)

The nuclei $^{57}$Mn and $^{60}$Fe serve as examples for another
source of differences between the shell model and FFN rates; the
latter being noticeably smaller than the shell model rates at low
temperatures/densities.  This is due to the use of different $Q_\beta$
values. FFN had to rely on the systematics available at the time,
while modern compilations indicate that the $Q_\beta$ values for these
nuclei are about 950 keV ($^{57}$Mn) and 1.7 MeV ($^{60}$Fe) smaller
than adopted by FFN. Obviously the too large $Q$ value suppressed the
electron capture at low temperatures/densities. With increasing
temperature/density the electron Fermi energy grows strongly reducing
the sensitivity to differences in the $Q$ value.

More generally, one expects that the electron capture rates become
less dependent on details of the GT strength distributions with
increasing electron Fermi energies. This explains why the deviations
between the FFN and shell model rates reduce with increasing
temperature/density. In this limit, the FFN rates should be still
slightly larger than the shell model rates due to the neglect of the
quenching of the total GT strength in \cite{FFN1,FFN2,FFN3}.

From the above discussion about the various contributions to the
rates, one might distinguish 3 different temperature/density regimes
along a stellar trajectory. At low $(T,\rho)$ specific low-lying
transitions can be quite important, supplementing the rate
contribution from the GT resonance. This is particularly the case for
electron capture, if the $Q$ value only allows capture of high-energy
electrons from the tail of the Fermi-Dirac distribution.  At
intermediate $(T,\rho)$ the rates are usually dominated by the strong
transitions involving the GT resonance. At high $(T,\rho)$ (when $E_e$
is large compared to $Q_{ij}$ for transitions to the GT centroid) the
rate becomes insensitive to the energy dependence of the GT
distribution and hence the rate depends only on the total GT strength.

Differences between the shell model results and the FFN assumptions in
the 3 ingredients (GT centroid energy, low-lying strength, $Q$ values)
lead also to differences in the $\beta^-$ decay rates, as is shown in
figure~\ref{fig:bd-ffn-sm}. Although the FFN $\beta$ decay rates are
usually somewhat larger than the shell model rates, no general picture
emerges in this comparison, as the rates are usually given by the sum
of low-lying transitions and contributions from the GT$_-$
distribution as well as of the back-resonances.  As is discussed in
\cite{Martinez99} the misplacement of the GT$_+$ centroid effects the
contribution of the back-resonances to the $\beta$-decay rates. In
particular, the back-resonances in odd-odd parents (the GT$_+$
centroids of even-even nuclei have been often placed at too high
energies by FFN) can be thermally excited more easily than assumed by
FFN resulting in slightly larger $\beta$-decay rates for these nuclei.
Examples are the odd-odd nuclei $^{60}$Mn and $^{60}$Cu. As FFN did
not consider the quenching of the GT$_+$ strength, the contribution of
the back-resonances is somewhat overestimated in FFN explaining the
slightly larger FFN $\beta$-decay rates for $^{54}$V or $^{54}$Mn.
Differences in the adopted Q-values effect mainly neutron-rich nuclei.
An example here is the $\beta$-decay of $^{57}$Cr, where FFN used the
Q-value of 6.56 MeV rather than 5.60~MeV.

The $pf$-shell nuclei discussed here have not too extreme $Y_e$ values
and are therefore important at the earlier stages of the presupernova
collapse involving low and intermediate ($T,\rho$) values. As in these
regimes the energy position of the GT centroid plays an essential
role, we will now investigate in details the differences between the
placement of this centroid in the FFN compilation with the shell model
rates (and the data).

In \cite{Martinez99} we have pointed out that there are systematic
differences between the placement of the GT centroid in the FFN
compilations and the shell model results (and data, if available).  It
has been noted that the differences apparently depend on the
pairing structure of the parent nucleus. For even-even parents, the
GT$_+$ centroid is calculated at slightly smaller energies than
assumed by FFN, while it has been noticed that for odd-$A$ and odd-odd
nuclei the centroid is at higher excitation energies than parametrized
by FFN. The consequences for the electron capture and $\beta$-decay
rates are obvious. If the kinematics is such that electron capture is
dominated by transitions to the GT resonance, the FFN rates should be
larger than the shell model rates as, for given temperature and
density, less electrons are available to capture to the centroid if it
resides at higher energies. This effect is strongest for odd-odd
nuclei where the differences between the FFN and shell model rates can
be larger than 2 orders of magnitude, i.e. for $^{60}$Co
(fig.~\ref{fig:ec-ffn-sm}) which is usually considered to be among the
most effective electron-capturing nuclei along the collapse trajectory
of a supernova.

Also for the $\beta$-decay rates the ratios systematically depend
on the pairing structure of the parent. Here, however, the shell model
rates are similar to the FFN rates for odd-odd parents (they are even
often slightly larger), while they are smaller for odd-A and even-even
nuclei. The reduction is usually largest for even-even nuclei.

The systematics assumed by FFN for the GT resonance energy in the
daughter is most easily illustrated for parent nuclei with only
$f_{7/2}$ protons and more than 28 neutrons (so that the $f_{7/2}$
orbital is blocked for GT$_+$ transitions). Then the single particle
contribution to the GT resonance energy is about 1.8 MeV, if the
neutron number is smaller $N\leq32$, or 0, if $N>32$. Thus one finds
$E_{GTR} \approx 3.8$ (2.0) MeV for even-even parent nuclei and 5.8
(4.0) MeV for odd-odd parents. For odd-$A$ parents one has
$E_{GTR}=3.8$ (2.0) MeV if the neutron number in the parent is even
and 5.8 (4.0) MeV, if it is odd. Here the numbers in parentheses refer
to parent nuclei with $N>32$ for which the $p_{3/2}$ neutron orbitals
are completely occupied.

\begin{figure}[ht]
  \begin{center}
    \leavevmode
    \epsfig{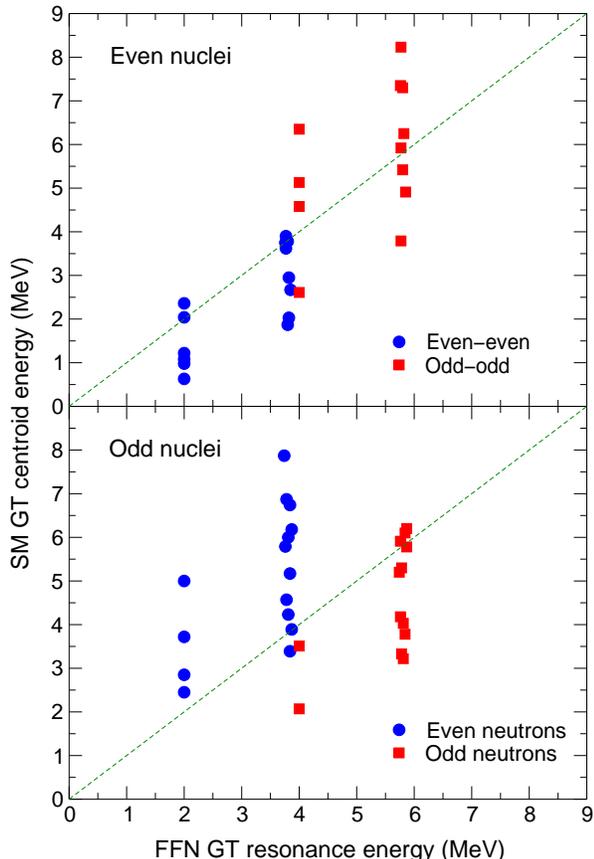}
    \caption{Comparison of the FFN GT$_+$ resonance energies with the
      shell model centroid energies for various nuclei with $Z \leq
      28$ and $N \ge 28 $.}
    \label{fig:ffnsm}
  \end{center}
\end{figure}

Figure~\ref{fig:ffnsm} compares the shell model GT centroids with the
FFN estimates of the GT resonance energy for about 45 nuclei with
proton numbers $Z\leq 28$ for which the above given systematic of the
FFN estimates apply. Indeed one clearly observes that the FFN GT
resonance energies cluster around 2, 4, and 6 MeV.  Clearly one
expects that the shell model GT centroid energies are more scattered
as the residual interaction fragments the GT strength and structure
effects will also effect the GT distributions. But a more striking
result is found if we compare the GT centroids for even-even, odd-$A$
and odd-odd parent nuclei separately. Here we find that, compared to
the shell model centroids, FFN places the GT resonance energy usually
at too high an energy for even-even parents and often at too low an
energy for odd-odd parents. The situation is more interesting and also
more telling for odd-$A$ parents. Note that all those nuclei, for
which FFN estimated the GT resonance energy by about 6 MeV, have
parent nuclei with an odd number of neutrons and thus require an
additional pairing energy in the FFN estimate. Compared to the shell
model GT centroids the FFN estimates for these nuclei are too high.
Then compare the nuclei with an even neutron number in the parent.
They do not acquire the additional pairing energy in the FFN estimate.
Compared to the shell model estimate, the FFN GT resonance energies
are usually too low.

\begin{table}[ht]
  \caption{Comparison of the experimental centroids of the GT
    resonance with the FFN GT resonance energies and the shell model
    centroids. The data are
    from~\protect\cite{gtdata1,gtdata2,gtdata3,gtdata4,gtdata5}. All
    energies are in MeV.}
  \label{tab:centro}
  \begin{center}
    \begin{tabular}{lccc}
      nucleus & data & FFN & shell model \\
      \hline
      $^{54}$Fe & $3.7 \pm 0.2$ &  3.80  & 3.78 \\
      $^{56}$Fe & $2.6 \pm 0.2$ &  3.78  & 2.60 \\
      $^{58}$Ni & $3.6 \pm 0.2$ &  3.76  & 3.75 \\
      $^{60}$Ni & $2.4 \pm 0.3$ &  2.00  & 2.88 \\
      $^{62}$Ni & $1.3 \pm 0.3$ &  2.00  & 1.78 \\
      $^{64}$Ni & $0.8 \pm 0.3$ &  2.00  & 0.50 \\
      $^{51}$V  & $4.8 \pm 0.2$ &  3.83  & 5.18 \\
      $^{55}$Mn & $4.1 \pm 0.3$ &  3.79  & 4.57 \\
      $^{59}$Co & $4.4 \pm 0.3$ &  2.00  & 5.05 \\
    \end{tabular}
  \end{center}
\end{table}

Thus it is obvious from this comparison that there is a different
dependence on the pairing structure of the parent ground state between
the FFN assumptions and the shell model results. But, which is
correct?  Experimentally the GT$_+$ strength distribution has been
studied for several even-even ($^{54,56}$Fe, $^{58,60,62,64}$Ni) and 3
odd-$A$ nuclei ($^{51}$V, $^{55}$Mn, $^{59}$Co) in this mass range. As
pointed out in \cite{Koonin94} the GT centroid for the even-even
parents are generally at lower excitation energies in the daughter
than for the odd-$A$ nuclei. If the experimental centroids are
compared to the FFN estimates one finds the same trend as discussed
above for the comparison with the shell model centroids. Here we have
calculated the experimental centroids from the measured GT
distributions upto 8 MeV. For $^{62,64}$Ni, however, we only consider
the peak in the GT distribution corresponding to the GT resonance. The
tail of the experimental distribution (see Fig. 1 in
\cite{Caurier99}), if real, is most likely due to states outside of
our present model space. As demonstrated in Table~\ref{tab:centro},
the FFN GT resonance energy is usually at too high an energy for
even-even parents, while it is at too low an energy for the 3 odd-$A$
nuclei, investigated experimentally. We note, however, that these 3
nuclei all have an even neutron number in the parent and just do not
allow to explore the assumptions which FFN made concerning the pairing
energy contribution to the GT resonance energy.  We stress that our
shell model centroids are generally in good agreement with the data,
as in fact the shell model calculations reproduce all measured GT$_+$
strength distributions quite satisfactory.

The differences in placement of the GT centroids explain the
differences between the FFN and shell model weak interaction rates.
As FFN assumes the GT resonance energy for odd-odd and odd-$A$ parent
nuclei with an even neutron number at lower energies than placed by
the shell model, electron capture to these strong transitions is
easier and hence the FFN electron capture rate is significantly larger
than the shell model rate for these nuclei. For electron capture on
even-even nuclei and for odd-$A$ nuclei with an odd neutron number,
FFN have generally placed the GT resonance energy at lower daughter
energies than predicted by the shell model (or the data). This effect
alone would make the FFN rates smaller than the shell model rates.
However, it is largely compensated by the fact that FFN did not
consider the quenching of the GT strength and by their consideration
of experimentally known transitions at low excitation energies. Taken
together, the FFN rates are also for these parent nuclei usually
somewhat smaller than the shell model rates, with the notable
exception of $^{56}$Ni \cite{Langanke98}. For reasons which will
become apparent in the next section, the FFN rates on very
neutron-rich odd-$A$ nuclei with odd neutron number are also often
smaller than the shell model rates.

Remembering the importance of the back-resonances for the $\beta^-$
decay rates, the effect of the different placements of the GT centroid
is obvious. In even-even nuclei and odd-$A$ nuclei with even neutron
number the shell model studies place the back-resonances at higher
excitation energies than assumed by FFN. Correspondingly, its thermal
population becomes less likely and hence the contribution of the
back-resonances to the $\beta^-$ decay rates decreases. On the
contrary, experimental data and the shell model calculations indicate
that the back-resonances reside actually at lower excitation energies
in odd-odd nuclei than assumed by FFN. Consequently, the contribution
of the back-resonances to the $\beta^-$ decay rate of odd-odd parent
nuclei should be larger than assumed in the FFN rates, which is indeed
the fact for several nuclei like $^{54,56}$Mn and $^{58}$Co. The
effect of the misplacement of the GT centroids on the $\beta^-$ rates
has already been discussed
in~\cite{Aufderheide93a,Aufderheide93b,Martinez99}.
 
In the next section we will argue why the systematics of the
 shell model GT centroids is correct. Furthermore, in connection with
the well-established systematics of the GT centroids for the GT$_-$
transitions, we will give a rather simple parametrization for these
quantities. Although these arguments still have to be delivered, we will close
this section pointing out that the present shell model rates are 
more reliable than the FFN ones and in fact should represent a fairly
accurate description of the nuclear structure problem required to derive
these rates. Then the weak interaction rates adopted in supernova
simulations should be revised. We will speculate briefly below about possible
consequences which the shell model rates might have for the presupernova
core collapse.

\section{Systematics of the GT centroid energies}

We begin our discussion by recalling that the systematics of the GT
centroid energy $E_{GT}$ is well understood for GT$_-$ transitions. In
fact, one finds in a very good approximation \cite{Nakayama}
\begin{equation}
E_{GT} - E_{IAS} =  a  +  b \cdot  \frac{(N-Z)}{A}.
\end{equation}
The constants $a,b$ can be derived by fit to measured GT$_-$
strengths~\cite{Nakayama}. 

\begin{figure}[ht]
  \begin{center}
    \leavevmode 
    \epsfig{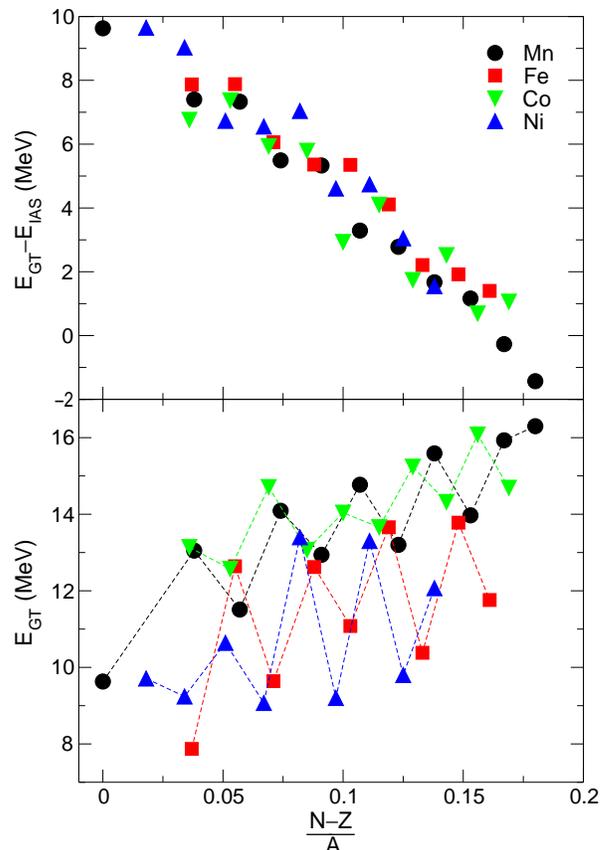}
    \caption{The shell model GT$_-$ centroid energy, relative to the
      energy of the isobaric analog state (upper panel) and to the
      daugther ground state (lower panel), for the Mn, Fe, Co, and Ni
      isotopes as function of $(N-Z)/A$. The pairing energy in the
      mass splitting between parent and daughter nuclei shifts the
      GT$_-$ centroid energy for odd-$A$ and odd-odd parents up in
      energy with respect to even-even parent nuclei.}
    \label{fig:centro}
  \end{center}
\end{figure}

In Fig.~\ref{fig:centro} we demonstrate that the centroids of our
shell model GT$_-$ strength distributions indeed exhibit this
$(N-Z)/A$ dependence. Our results show a spread of about 1 MeV which
can be interpreted as the uncertainty introduced by nuclear structure
effects. In passing we note that for nuclei with large neutron excess
the GT$_-$ centroid energy can be below the IAS energy.  As the figure
is compiled for the isotope chains with $Z=25-28$ it comprises all 3
different types of parent nuclei: even-even, odd-odd and odd-$A$.
Clearly no systematic dependence on the pairing structure is found.
Obviously such a dependence shows up if the GT centroid energies are
plotted as measured relatively to the daughter ground state energy
(lower panel of figure~\ref{fig:centro}). It is trivially introduced
by the differences in pairing energy between the various parent and
daughter nuclei. As a consequence GT$_-$ centroids form now 3
distinguished bands: one for even-even nuclei, one for odd-$A$ nuclei,
and one for odd-odd nuclei the latter two shifted up in energy by
about twice or four times the pairing energy, respectively.

As we will show now the same behavior is found for the centroids of
the GT$_+$ strength distributions. Before we do so, however, it is
useful to recall the origin of the $(N-Z)/A$ dependence of the GT$_-$
centroids.  Here we will follow the nice discussion given by Bertsch
and Esbensen~\cite{Esbensen}.  Then the average excitation energy of
the GT operator is given by
\begin{equation}
E_{GT} = \int_0^\infty dE E S(E) = \frac{\langle \bbox{\sigma}
  \bbox{t}_\mp [H, \bbox{\sigma} \bbox{t}_\pm] \rangle}{\langle
  \bbox{\sigma} \bbox{t}_\mp \bbox{\sigma} \bbox{t}_\pm \rangle}
\end{equation}
where the expression is valid for both GT$_-$ and GT$_+$ operators.
Several pieces of the Hamiltonian do not commute with the GT operator.
While the most of these contributions are cancelled in building the
difference with the IAS energy, the $v_{\sigma \tau} \bbox{\sigma}_1
\cdot \bbox{\sigma}_2 \bbox{\tau}_1 \cdot \bbox{\tau}_2$ residual
interaction gives rise to an energy shift that in the Tamm-Dancoff
approximation is given by~\cite{Esbensen}
\begin{equation}
  \label{eq:deltast}
  \Delta E_{\sigma\tau} = \frac{\langle v_{\sigma \tau} \rho \rangle 2
    S_\beta}{3A}
\end{equation}
where ${\it v}_{\sigma \tau} \rho$ is the product of the integrated
strength and the averaged ground state density.  (In building the
difference with the IAS energy a similar contribution arising from the
$v_\tau \bbox{\tau}_1 \cdot \bbox{\tau}_2$ interaction has to be
subtracted.) $S_\beta$ is the total GT strength, which for the GT$_-$
operator and neutron-rich nuclei, can be approximated by the Ikeda
sumrule, $S_{\beta^-} = 3 (N-Z)$. Upon substitution
into~(\ref{eq:deltast}), one finds the desired $(N-Z)/A$ dependence.

For the GT$_+$ operator one derives at the same expression for $\Delta
E_{\sigma \tau}$, however, now considering $S_{\beta^+}$. If one
measures the GT$_+$ centroid from the parent ground state, the
contribution from the other terms of the Hamiltonian are largely
cancelled and one has upto a constant (reflecting for example the
spin-orbit splitting) $E_{GT} = \text{const} + E_{\sigma \tau}$.  The
problem just reduces to find an appropriate parametrization of the
total GT$_+$ strength. As has been pointed out in
Ref.~\cite{Koonin94}, the presently available data for $pf$ shell
nuclei suggest a scaling of the total GT$_+$ strength like
\begin{equation}
S_{\beta^+} = a Z_v \cdot (20-N_v)
\end{equation}
where $Z_v, N_v$ are the number of valence protons and neutrons in the
$pf$ shell, respectively. This dependence corresponds to a generalized
BCS model with pure proton and neutron pairing \cite{Engel97}.

\begin{figure}[ht]
  \begin{center}
    \leavevmode
    \epsfig{file=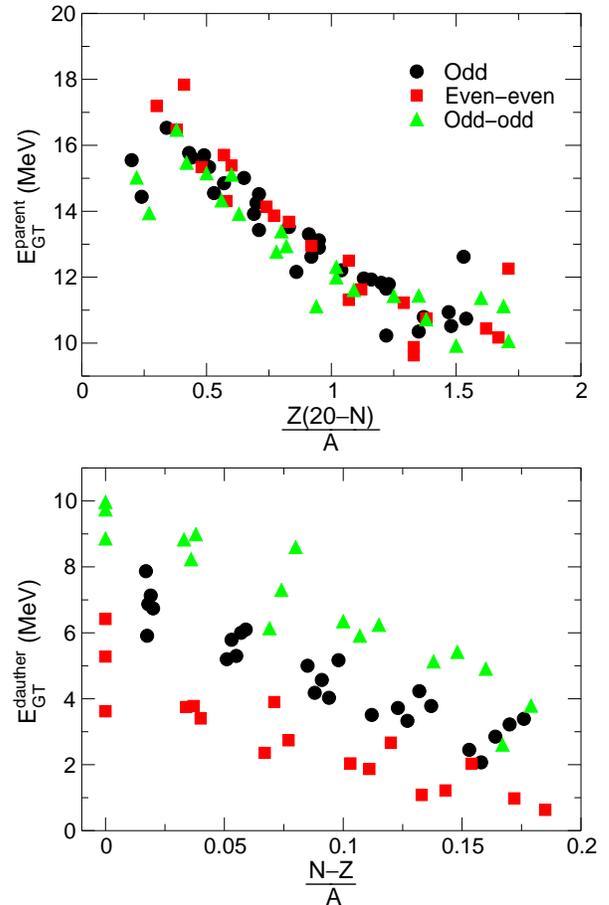,width=0.9\columnwidth}
    \caption{Upper panel: shell model GT$_+$ centroid energy, relative to the
      parent ground state energy, for various nuclei in
      the mass range $A=50-60$ as function of $Z_v (20-N_v)/A$. Lower
      panel: shell model GT$_+$ centroid energy, relative to the
      daughter ground state energy, for the same nuclei as function of
      $(N-Z)/A$. The pairing energy in the mass splitting between
      parent and daughter nuclei shifts the GT$_+$ centroid energy for
      odd-$A$ and odd-odd parents up in energy with respect to
      even-even parent nuclei.}
    \label{fig:egt}
  \end{center}
\end{figure}

We have calculated the $GT_+$ centroids for about 75 nuclei in the
$pf$ shell and if we plot these energies with respect to the parent
ground state energies, they closely follow the $Z_v \cdot (20-N_v) /A$
rule.  Importantly, like for the case of the GT$_-$ centroids, no
dependence on the pairing structure of the parent nucleus is observed.
However, such a dependence, as we have seen above for the GT$_-$
centroids, is introduced if one measures the GT centroids with respect
to the daughter ground state energies. This is again demonstrated in
Fig.~\ref{fig:egt}. Here we have chosen to plot the centroid energies
as function of $(N-Z)$ which is the dominating dependence in the
parent-daughter mass splitting and is stronger than the $Z_v (20-N_v)$
dependence of $S_{\beta^+}$.  Like for GT$_-$, the centroids now group
according to the pairing structure of the parent ground state. Again
the centroids are lowest for even-even parents and the centroids for
the odd-$A$ and odd-odd parents are shifted up in energy by about 3
and 6 MeV, resp.

We note that the same pairing shifts of the $S_{\beta^+}$ strength has
already been suggested by Hansen in a general discussion of the
$\beta$ strength in nuclei~\cite{Hansen}.

\begin{figure}[ht]
  \begin{center}
    \leavevmode
    \epsfig{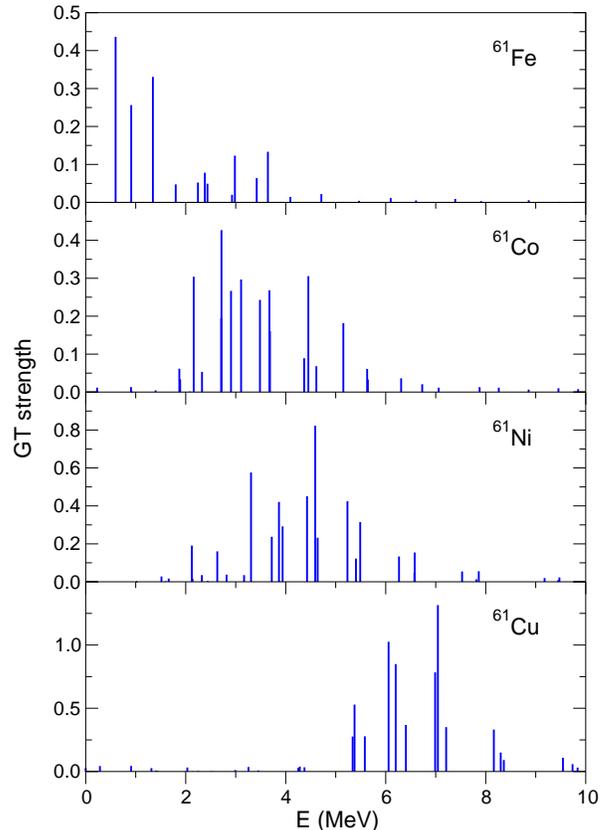}
    \caption{The shell model GT$_+$ distributions for the odd $A=61$
      isotones as function of excitation energy in the daughter
      nucleus. Note the different scales used for the various nuclei.}
    \label{fig:oddA}
  \end{center}
\end{figure}

The dependence of the GT$_+$ distribution on the proton and neutron
number is nicely visualized in fig.~\ref{fig:oddA} for the odd-$A=61$
isotone chain. For comparison we note that the independent particle
model, as assumed by FFN, places the GT$_+$ centroids at excitation
energies of 4 MeV for $^{61}$Fe and $^{61}$Ni (the daughter nuclei
have an even neutron number) and at 2 MeV for $^{61}$Co. $^{61}$Cu has
9 valence protons, thus allowing also for a $p_{3/2}$ proton being
changed into a $p_{1/2}$ neutron. Relatedly, the GT$_+$ centroid is
shifted slightly to 2.1 MeV.  As has already been visible in
fig.~\ref{fig:ffnsm}, the shell model shows quite a different
dependence of the GT$_+$ distributions. At first, we observe that in
all cases the strength is concentrated in an energy region of about
3-4~MeV width.  Further, the centroid of this region decreases with
increasing neutron excess. However, we stress that this decrease is
basically due to the $(N-Z)$ dependence of the mass difference between
parent and daughter nucleus, as the GT$_+$ centroid energy increases
slightly with $(N-Z)$ if measured with respect to the parent ground
state. The figure clearly shows no distinct dependence of the GT$_+$
centroid energy on the neutron pair configuration. We find the shell
model centroids at excitation energies of 2.1 MeV ($^{61}$Fe), 3.7 MeV
($^{61}$Co), 4.7 MeV ($^{61}$Ni), and 6.7 MeV ($^{61}$Cu).  Clearly
the total GT$_+$ strength decreases with increasing neutron excess in
the isotone chain due to the decreasing number of valence protons and
the increasing Pauli blocking of the neutrons.

Finally we remark that the figure also shows that the thermal
excitation of the strong backresonance transitions becomes easier with
increasing neutron excess, as these transitions move to lower
excitations energies.

\section{Conclusions}

Bethe and collaborators \cite{BBAL} had, more than two decades ago,
focussed the attention on the importance played by the weak
interaction in a supernova collapse.  Subsequently about twenty years
ago Fuller, Fowler and Newman (FFN) \cite{FFN1,FFN2,FFN3} outlined in
their seminal work the theory to calculate stellar weak interaction
rates. After this pioneering step the problem had been reduced to
solve the related nuclear structure physics. While FFN understood the
relevant physics correctly, due to lack of data and computational
resources they were forced to estimate the relevant weak interaction
rates on nuclei in the mass range $A=45-60$ phenomenologically. Over
the years experimental findings about the fragmentation and
positioning of the GT strength in these nuclei indicated the need for
refinements of the rates and it became
obvious~\cite{Aufderheide93a,Aufderheide93b} that the interacting
shell model is the method of choice for this endeavour. First steps
towards this goal have been undertaken using the shell model Monte
Carlo technique \cite{Koonin}, but it became clear that shell model
diagonalization is the better suited tool to calculate reliable
stellar rates. Impressive progress in both shell model programming (in
particular due to the work by E.  Caurier) and hardware development
allows now for virtually converged calculations of the GT strength in
$pf$ shell nuclei, as they play a fundamental role during the
presupernova collapse. In fact, Caurier {\it et al.} have demonstrated
\cite{Caurier99} that shell model diagonalization is able to reproduce
all measured GT$_+$ and GT$_-$ strength distributions on nuclei around
$A\approx 60$ and simultaneously describe the spectra and lifetimes of
these nuclei also sufficiently well. As this is the relevant input to
reliably calculate stellar weak interaction rates, it has been
concluded that the shell model diagonalization approach of Ref.
\cite{Caurier99} has the necessary predictative power to calculate the
rates. In this manuscript we have followed this conclusion and have
derived stellar weak interaction rates based on state-of-the-art shell
model diagonalization. The calculation has been performed for more
than 100 nuclei in the mass range $A=45-65$ and covers the temperature
and density regime expected in supernova physics. The rates have been
compiled in a file using the same format as is customary for the FFN
rates. The electronic version of the file can be received from the
authors upon request. The files are also available in the effective
rates formalism as derived in \cite{FFN4}.

Differences between the shell model and the FFN rates are usually
related to differences in the placement of the GT resonance energies.
In analogy to the wellknown systematics of the GT$_-$ centroid
energies we have derived a similar systematics for the GT$_+$
resonance energies.  In particular we have shown that the GT$_+$
centroid energies, if measured with respect to the parent ground state
energy, are not dependent on the pairing structure of the parent
nucleus. If the centroids are, however, measured with respect to the
daughter ground state energy, the wellknown pairing energy
contributions to the mass splitting between parent and daughter nuclei
enters and the positions of the GT centroids are pushed up in energy
for odd-$A$ and odd-odd parents by twice and four-times the pairing
energy, respectively. This systematic has not quite been considered in
previous estimates of the weak interaction
rates~\cite{FFN2,FFN3,Aufderheide94,Sutaria}.

\begin{figure}[ht]
  \begin{center}
    \leavevmode
    \epsfig{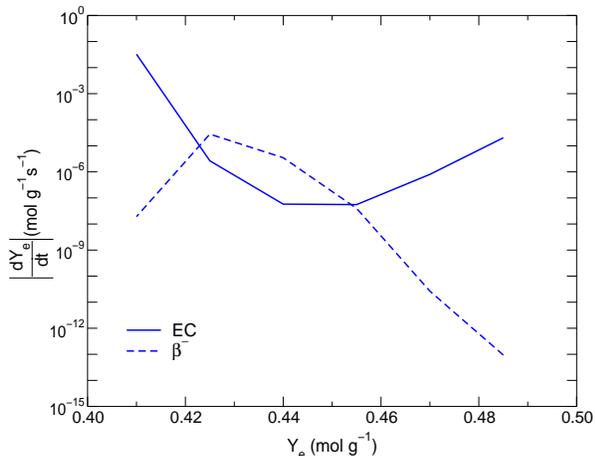}
    \caption{Change in the total electron  capture and $\beta$-decay
      rates as a function of electron-to-baryon ratio $Y_e$ during a
      supernova collapse.  The respective values for the temperature
      and density were taken from the stellar trajectory given
      in~\protect\cite{Aufderheide94}.}
    \label{fig20}
  \end{center}
\end{figure}

The shell model rates are usually smaller than the FFN rates. This is
particularly the case for the electron capture rate on odd-odd nuclei
and to a lesser extent on odd-$A$ nuclei. This can be examplified for
the nuclei $^{55}$Co, $^{57}$Co, $^{54}$Mn, $^{60}$Co and $^{58}$Mn
which the compilation of \cite{Aufderheide94} subsequently ranks as
the most effective electron capturing nuclei in the density regime
$\rho=10^7-10^{10}$ g/cm$^3$.  For these nuclei the shell model rates
are smaller than the FFN rates by factors 39, 12, 10, 346, and 19,
respectively, where the comparison is made for those temperatures and
densities where Ref. \cite{Aufderheide94} lists the respective nucleus
as most important. Already this comparison suggests that stellar
electron capture rates are noticeably smaller than previously assumed.
As speculated in \cite{Langanke99}, due to the smaller electron
capture rates the core should radiate less energy away by neutrino
emission, keeping the core on a trajectory with higher temperature and
entropy.

Electron capture has to compete with $\beta$ decay in the stellar
environment. For even-even and odd-$A$ nuclei, the shell model
$\beta$-decay rates are generally also noticeably smaller than the FFN
rates. But for odd-odd nuclei shell model and FFN rates are about the
same. As the $\beta$ decay of odd-odd nuclei is expected to contribute
significantly to the total $\beta$ decay rate during a collapse, the
total $\beta$-decay rate should change less in supernova simulations
than the electron capture rate, if the FFN compilation is replaced by
the shell model results.

Obviously collapse calculations which use the shell model stellar weak
interaction rates are desired and those studies are already initiated.
Definite conclusions have to wait for the results of these
simulations, but we can here update one interesting finding put
forward in \cite{Aufderheide96,Martinez99}. These authors argued that
during the collapse the $\beta$ decay rate will exceed the electron
capture rate for a certain range of electron-to-baryon ratios $Y_e$.
As a consequence the core can radiate energy away, (as the neutrinos
can still leave the star without interaction) without lowering the
$Y_e$ value.  This should have some interesting consequences for the
size of the homologous core.

In \cite{Martinez99} the regime in which the $\beta$-decay rate
exceeds the electron capture rate has been estimated on the basis of
shell model rates for a limited set of nuclei. Here we update this
comparison by adopting the full set of shell model rates.  To do so,
we follow Ref. \cite{Aufderheide96} and define the change of $Y_e$ due
to $\beta$ decay (this increases the charge by one unit) and electron
capture (which reduces the charge by one unit) as
\begin{equation}
{\dot Y}_e^{ec(\beta)} = \frac{dY_e^{ec(\beta)}}{dt}=-(+) \sum_k
\frac{X_k}{A_k} \lambda_k^{ec(\beta)}
\end{equation}
where the sum runs over all nuclear species present in the core.  and
$\lambda_k^{ec}$ and $\lambda_k^{\beta}$ are the electron capture and
$\beta$ decay rates of nucleus $k$.  The mass fraction is given by
nuclear statistical equilibrium \cite{Aufderheide94}.

As in \cite{Martinez99} we will follow the stellar trajectory as given
in Ref.  \cite{Aufderheide96}, although this is expected to change
somewhat if the FFN rates are replaced by the shell model rates in the
collapse simulation. Fig.  \ref{fig20} compares ${\dot Y}_e^{ec}$ and
${\dot Y}_e^{\beta}$ along the stellar trajectory where $Y_e$ reduces
here with time. Confirming the results of \cite{Martinez99} the full
set of shell model rates also reveals that the $\beta$ decay rates are
larger than the electron capture rates for $Y_e=0.42-0.455$.  This
might have important consequences for the core collapse possibly
leading to cooler cores and larger $Y_e$ values at the formation of
the homologuous core.

Thielemann and collaborators \cite{Thielemann,Brachwitz} have reported
first attempts to explore the role of the shell model electron capture
rates in type Ia supernovae. They find that the composition of the
matter in the center is less neutron-rich than previously assumed. If
the FFN rates are replaced by the shell model ones the overproduction
of the neutron-rich Cr, Ti, and Fe isotopes which has been encountered
in previous type Ia simulations with otherwise the same physics input
\cite{Iwamoto}, is removed. As the present shell model rates likely
reduce the uncertainties related to the stellar electron capture
rates, it is expected that type Ia simulations with the new shell
model electron capture rates will serve as strict tests for the models
and their parametrizations.  Such comprehensive studies are in
progress.

\acknowledgements

It is a pleasure to thank E. Caurier, F. Nowacki and A. Poves who have
supplied us with the effective interaction used in the our shell model
calculations. We also like to thank D. Arnett, G.E. Brown, G.M. Fuller, 
F.-K. Thielemann
and S.E. Woosley for valuable discussions about the astrophysical aspects of
the weak interaction rates.
This work was supported in part by the Danish Research Council.
Computational resources were provided by the Center for Advanced
Computational Research at Caltech.

\end{document}